\documentclass[aps,reprint,twocolumn,superscriptaddress]{revtex4}
\usepackage{amsfonts}
\usepackage{amsmath}
\usepackage{amssymb}
\usepackage{graphicx}
\usepackage{bm}
\usepackage{color}
\usepackage{mathrsfs}
\usepackage{mathtools}
\usepackage[colorlinks,bookmarks=true,citecolor=blue,linkcolor=red,urlcolor=blue]{hyperref}
\usepackage{multirow}
\usepackage{makecell}
\usepackage{appendix}
\usepackage{float}
\usepackage{url}
\usepackage{verbatim} 
\newtheorem{theorem}{Theorem}
\begin{document}
\title{A $\bm{k}\cdot\bm{p}$ effective Hamiltonian generator}
\author{Yi Jiang}
\affiliation{Beijing National Laboratory for Condensed Matter Physics and Institute of Physics, Chinese Academy of Sciences, Beijing 100190, China}
\affiliation{University of Chinese Academy of Sciences, Beijing 100049, China}
\author{Zhong Fang}
\affiliation{Beijing National Laboratory for Condensed Matter Physics and Institute of Physics, Chinese Academy of Sciences, Beijing 100190, China}
\author{Chen Fang}
\email{cfang@iphy.ac.cn}
\affiliation{Beijing National Laboratory for Condensed Matter Physics and Institute of Physics, Chinese Academy of Sciences, Beijing 100190, China}
\affiliation{Songshan Lake Materials Laboratory, Dongguan, Guangdong 523808, China}
\affiliation{Kavli Institute for Theoretical Sciences, Chinese Academy of Sciences, Beijing 100190, China}

\begin{abstract}
$\bm{k}\cdot\bm{p}$ effective Hamiltonian is important for theoretical analysis in condensed matter physics. Based on the \textit{kdotp-symmetry} package, we develop an upgraded package named \textit{kdotp-generator}. This generator takes in arbitrary magnetic symmetries with their representations and returns symmetry-allowed $\bm{k}\cdot\bm{p}$ Hamiltonians. Using this package, we calculate $\bm{k}\cdot\bm{p}$ Hamiltonians for irreducible corepresentations in 1651 magnetic space groups up to the third order, and their linear coupling to external fields including the electromagnetic field and the strain tensor. 
We hope that the package will facilitate related research in the future.
\end{abstract}
\maketitle

\section{Introduction}
Condensed matter systems usually possess great complexities, and it is important to make simplifications, which leads to the effective Hamiltonian. The $\bm{k}\cdot\bm{p}$ effective Hamiltonians (or $\bm{k}\cdot\bm{p}$ models for short) were invented in the middle twentieth century to describe the dispersion of metals and semiconductors\cite{bardeen1938improved, PhysRev.58.633, PhysRev.97.869, PhysRev.98.368, kane1956energy, PhysRev.102.1030,  kane1966k}, derived using perturbation theory. They are intended to model the dynamics of quasiparticles near specific momenta. 
With $\bm{k}\cdot\bm{p}$ models, it becomes possible to analytically calculate the physical properties of materials, including nontrivial topologies and novel responses to various external fields\cite{hasan2010colloquium, qi2011topological, chiu2016classification, armitage2018weyl, zhang2009topological, yu2010quantized, xu2011chern, wan2011topological, burkov2011topological, young2012dirac, wang2012dirac, hsieh2012topological, wang2013three, weng2015weyl, wang2016hourglass, ruan2016symmetry, PhysRevLett.116.226801, soluyanov2015type, bradlyn2016beyond, gresch2017hidden, yang2020unlocking, yang2021symmetry}.

Unlike early perturbative Hamiltonians constructed from atomic wavefunctions, 
the most general form of a $\bm{k}\cdot\bm{p}$ model can be determined by the symmetry group and the representations of energy bands\cite{PhysRev.102.1030, voon2009kp}. In literature, the symmetry-allowed $\bm{k}\cdot\bm{p}$ models are usually derived case-by-case for (magnetic) space groups. An automated package \textit{kdotp-symmetry} for calculating $\bm{k}\cdot\bm{p}$ models was developed by D. Gresch in 2018\cite{gresch2018identifying} at \href{http://z2pack.ethz.ch/kdotp-symmetry/}{http://z2pack.ethz.ch/kdotp-symmetry/}. This package can take in both unitary and anti-unitary symmetry operations and their representation matrices (either reducible or irreducible), and return all compatible, linearly independent $\bm{k}\cdot\bm{p}$ models.

In this work, based on the \textit{kdotp-symmetry} package, we make a few improvements and develop a modified package \textit{kdotp-generator} at \href{https://github.com/yjiang-iop/kdotp-generator}{https://github.com/yjiang-iop/kdotp-generator}. Besides $\bm{k}\cdot\bm{p}$ Hamiltonians, our package can also calculate the symmetry-allowed Hamiltonians coupled to external fields including the electromagnetic field $\bm{E},\bm{B}$ and the strain tensor $\epsilon_{\mu\nu}$. The output Hamiltonians are decomposed symmetrically using linear representations. 
We pre-compute $\bm{k}\cdot\bm{p}$ Hamiltonians for irreducible corepresentations (coirreps) in 1651 magnetic space groups (MSGs)\cite{bradley1968magnetic} of order $\le 3$, and their linear coupling to external fields. In the following, we will first review the algorithms for deriving $\bm{k}\cdot\bm{p}$ models, and then introduce our package and give some examples.

\section{Algorithm}
The $\bm{k}\cdot\bm{p}$ effective Hamiltonian $H(\bm{k})$ is the asymptotic expansion of the system's Hamiltonian near a chosen high-symmetry momenta $\bm{k}_0$, where we use $\bm{k}$ to denote the small deviation from $\bm{k}_0$ ($\bm{k}_0$ is omitted in $H(\bm{k})$ for simplicity). Assume first the little group $G$ of $\bm{k}_0$ has only unitary spacial symmetries, and $H(\bm{k})$ belongs to a n-dimensional representation $D$ (either reducible or irreducible).
The Hamiltonian must satisfy the symmetry constraint equation
\begin{equation}
\begin{aligned}
H(\bm{k})=D(g)H(g^{-1}\bm{k})D^{-1}(g)
\end{aligned}
\label{constraint}
\end{equation}
 At $\bm{k}=\bm{0}$, Eq.(\ref{constraint}) reduces to the familiar commutation relation $[H(\bm{0}), D(g)]=0$.

The $\bm{k}\cdot\bm{p}$ model can be expanded using $\bm{k}$-monomials and Hermitian matrices. 
There are $N=n^2$ linearly independent n-dimensional Hermitian matrices $\vec{X}=\{X_i, i=1,...,n^2\}$ and $M=\frac{1}{2}(m+1)(m+2)$ $m$-th order $\bm{k}$-monomial $\vec{f}(\bm{k})=\{k_x^ik_y^jk_z^k | i,j,k\in \mathbb{Z}_{\ge 0}, i+j+k=m \}$. 
The vector spaces they spanned, i.e., $\{\sum_l c_l X_l|c_l\in\mathbb{R}\}$ and $\{\sum_l c_l f_l(\bm{k})|c_l\in\mathbb{R}\}$, 
are closed under the group actions of $g\in G$ defined by $\hat{G}_g$ and $\hat{F}_g$:
\begin{equation}
\begin{aligned}
\hat{G}_g X_l &\coloneqq D(g)X_l D^{-1}(g)=\sum_{l'} X_{l'}M_{l'l}(\hat{G}_g)\\
\hat{F}_gf_l(\bm{k})&\coloneqq f_l(g^{-1}\bm{k})=\sum_{l'}f_{l'}(\bm{k})N_{l'l}(\hat{F}_g)
\end{aligned}
\end{equation}
where $M(\hat{G}_g)_{N\times N}$ and $N(\hat{F}_g)_{M\times M}$ are the (real) representation matrices of $G$ in these two vector spaces, which are generally reducible.

The $\bm{k}\cdot\bm{p}$ Hamiltonian is defined on the direct product space $\vec{f}(\bm{k}) \otimes \vec{X}$, and the symmetry constraint Eq.\ref{constraint} is equivalent to
\begin{equation}
(\hat{F}_g\otimes \hat{G}_g)H(\bm{k})=H(\bm{k}),\ \ \forall g\in G
\label{operator_constraint}
\end{equation}
which means $H(\bm{k})$ belongs to the trivial representation of $\hat{F}_g\otimes \hat{G}_g$.

The little group of $\bm{k_0}$ may also contain anti-unitary symmetry $h=Tg_0$, with $T$ being the time-reversal symmetry (TRS) and $g_0$ a spacial symmetry. 
The representations of magnetic little groups can be constructed from the irreducible representations (irreps) of unitary subgroup, which are called irreducible corepresentations (coirreps). For a $\bm{k}\cdot\bm{p}$ Hamiltonian with magnetic little group $M=G+hG$ and belonging to a corepresentation $D$ of $M$, it must satisfy the additional symmetry constraint cast by the anti-unitary symmetry $h$:
\begin{equation}
\begin{aligned}
H(\bm{k})&=U(Tg_0)H^*(-g_0^{-1}\bm{k}) U^{-1}(Tg_0)\\
\end{aligned}
\label{au_constraint}
\end{equation}
where we have used $D(Tg_0)=U(Tg_0)\hat{K}$, with $U(Tg_0)$ being the representation matrix and $\hat{K}$ the complex conjugation. The group actions of $h$ on Hermitian matrices and $\bm{k}$-monomials are also modified:
\begin{equation}
\begin{aligned}
\hat{G}_h X_l&=U(Tg_0) X_l^* U^{-1}(Tg_0) = \sum_{l'}X_{l'} M_{l'l}(\hat{G}_h)\\
\hat{F}_h f_l(\bm{k}) &=
f_l(-g_0^{-1}\bm{k}) = \sum_{l'}f_{l'}(\bm{k}) N_{l'l}(\hat{F}_h)
\label{au_action}
\end{aligned}
\end{equation}

The $\bm{k}\cdot\bm{p}$ Hamiltonian may also depend on external fields like the electromagnetic field $\bm{E},\bm{B}$ and the strain tensor $\epsilon_{\mu\nu}$, which satisfies similar symmetry constraint:
\begin{equation}
\begin{aligned}
&H(\bm{k},\bm{E},\bm{B},\epsilon)\\
&=D(g)H(g^{-1}\bm{k}, g^{-1}\bm{E}, g^{-1}\bm{B}, g^{-1}\epsilon g)D^{-1}(g)
\end{aligned}
\end{equation}

These external fields have different space-time transformation properties. For proper rotations $R\in \text{SO}(3)$, space-inversion $P$, and time-reversal $T$, they transform as:
\begin{table}[ht]
	\centering
	\begin{tabular}{c|c|c|c}
		\hline\hline
		Tensor & $P$ & $T$ & $R$ \\\hline
		Polar vector $\bm{k}$ & $-\bm{k}$ & $-\bm{k}$ & $R\bm{k}$ \\\hline
		Polar vector $\bm{E}$ & $-\bm{E}$ & $\bm{E}$ & $R\bm{E}$ \\\hline
		Axial vector $\bm{B}$ & $\bm{B}$ & -$\bm{B}$ & $R\bm{B}$ \\\hline
		2nd order tensor $\epsilon_{\mu\nu}$ & $\epsilon_{\mu\nu}$ & 
		$\epsilon_{\mu\nu}$ & $R_{\mu\mu^\prime}R_{\nu\nu^\prime}\epsilon_{\mu^{\prime}\nu^{\prime}}$ 
		\\\hline\hline
	\end{tabular}
	\caption{\label{tensor} The transformation properties of $\bm{k}$ and external fields $\bm{E},\bm{B},\epsilon_{\mu\nu}$.}
\end{table}

Note that $\epsilon_{\mu\nu}$ can also be seen as two independent polar vectors $\epsilon_{\mu}\epsilon_{\nu}$ which transform as $\epsilon_{\mu}^\prime\epsilon_{\nu}^\prime=
R_{\mu\mu^\prime}\epsilon_{\mu^{\prime}}
R_{\nu\nu^\prime}\epsilon_{\nu^{\prime}}$.

In the following, we introduce two algorithms for computing symmetry-allowed $\bm{k}\cdot\bm{p}$ models, in which the external fields can be treated in equal footing as $\bm{k}$.

\begin{figure*}
	\centering
	\includegraphics[width=1\textwidth]{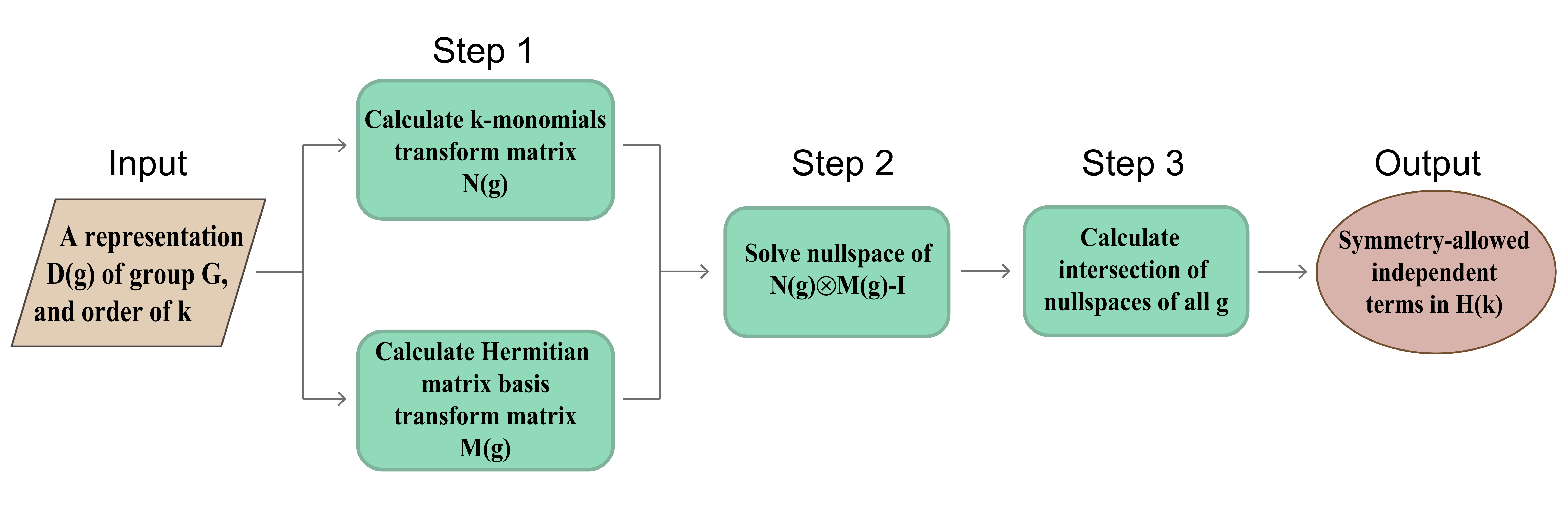}
	\caption{\label{flowchart}The flowchart of direct-product decomposition method for computing $\bm{k}\cdot\bm{p}$ models.}
\end{figure*}

\subsection{The direct-product decomposition method}
The first algorithm, which we call the ``direct-product decomposition method'', was introduced by D. Gresch in the $\emph{kdotp-symmetry}$ package\cite{gresch2018identifying}. In this method, the constraint Eq.\ref{operator_constraint} is reformulated as 
\begin{equation}
H(\bm{k})\in \bigcap_{g \in M} \operatorname{Eig}\left(\hat{F}_{g} \otimes \hat{G}_{g}, 1\right)
\end{equation}
where $\operatorname{Eig}\left(\hat{F}_{g} \otimes \hat{G}_{g}, 1\right)$ denotes the eigenspace of operator $\hat{F}_{g} \otimes \hat{G}_{g}$ with eigenvalue 1.

To find all $\bm{k}\cdot\bm{p}$ models that satisfy the symmetry constraints is equivalent to solve the nullspace of
\begin{equation}
N(\hat{F}_g)\otimes M(\hat{G}_g) - \mathbb{I}_{MN\times MN}
\end{equation}
for each $g\in M$, and then calculate the intersection of the nullspace of all $g$.
Each independent basis of the nullspace corresponds to one linearly independent $\bm{k}\cdot\bm{p}$ Hamiltonian. 
The Hermiticity of the computed $\bm{k}\cdot\bm{p}$ models is ensured, as both $M(\hat{G}_g)$ and $N(\hat{F}_g)$ are real matrices which have real null vectors.
This algorithm is summarized in Fig.\ref{flowchart}.

\subsection{The irrep-matching method}
The second algorithm, which we call the ``irrep-matching method'', dates back to Ref.\cite{PhysRev.102.1030}. It makes use of the linear coirreps of the magnetic little group to classify the Hermitian matrices and $\bm{k}$-monomials into symmetric bases, and then matches the bases of the same coirrep. 

Under the action of $g\in M$, the Hermitian matrices and $\bm{k}$-monomials can be rearranged such that they form the bases of linear coirreps of $M$, i.e.
\begin{equation}
\begin{aligned}
	\vec{X}&\simeq\oplus_{\kappa,\mu}\vec{X}^{\kappa,\mu},\ \
	\vec{f}(\bm{k})\simeq\oplus_{\kappa,\nu}\vec{f}^{\kappa,\nu}(\bm{k})
\end{aligned}
\end{equation}
where $\kappa$ denotes the $\kappa$-th linear coirrep $D^\kappa$ of $M$, $\mu, \nu$ denote different sets of bases of $D^\kappa$ (more than one set of bases could exist), and 
\begin{equation}
X_l^{\kappa,\mu}=\sum_{l^\prime}X_{l^\prime}U^{\kappa,\mu}_{l^\prime l},\ \  f_l^{\kappa,\nu}(\bm{k}) =\sum_{l^\prime}f_{l^\prime}(\bm{k})V^{\kappa,\nu}_{l^\prime l}
\end{equation}
are symmetric bases with $U^{\kappa,\mu},V^{\kappa,\nu}$ being the unitary similarity transformation matrices. These symmetric bases transform as
\begin{equation}
\begin{aligned}
\hat{G}_gX_l^{\kappa,\mu}&=D(g) X_l^{\kappa,\mu} D^{-1}(g) = \sum_{l^\prime}X_{l^\prime}^{\kappa,\mu}D^\kappa_{l^\prime l}(g)\\
\hat{F}_gf_l^{\kappa,\nu}&=f_l^{\kappa,\nu}(g^{-1}\bm{k}) =\sum_{l^\prime} f_{l^\prime}^{\kappa,\nu}(\bm{k})D^\kappa_{l^\prime l}(g)
\end{aligned}
\end{equation}
which can be written into more compact form:
\begin{equation}
	\hat{G}_g\vec{X}^{\kappa,\mu}=\vec{X}^{\kappa,\mu}D^\kappa(g),\ \
	\hat{F}_g\vec{f}^{\kappa,\nu}=\vec{f}^{\kappa,\nu}D^\kappa(g)
\end{equation}
Notice that $D^\kappa$ here are linear coirreps of $M$, as the group action $\hat{G}_g$ and $\hat{F}_g$ depend solely on the $\text{O}(3)$ rotation part of $g$, but independent of the translation and $\text{SU}(2)$ part.

With these symmetric bases, $H(\bm{k})$ can be expanded as 
\begin{equation}
H(\bm{k})=\sum_{\kappa,\mu,\nu} a_{\kappa \mu\nu} \vec{X}^{\kappa,\mu}\vec{f}^{\kappa, \nu\dagger}(\bm{k})
\label{kp_hamiltonian}
\end{equation}
which is a linear combination of $\vec{X}^{\kappa,\mu}$ times $\vec{f}^{\kappa,\nu}(\bm{k})$ that belong to the same linear coirrep $D^\kappa$, and $a_{\kappa\mu\nu}$ are real parameters. We claim that the Hamiltonian thus expanded processes $M$ symmetry and belongs to the corepresentation D, as long as each $D^\kappa$ is a unitary linear coirrep. The proof for unitary symmetries is as follows:
\begin{equation}
\begin{aligned}
&D(g)H(g^{-1}\bm{k})D^{-1}(g) \\
&=\sum_{\kappa,\mu,\nu}a_{\kappa \mu\nu} \sum_l D(g)X_l^{\kappa,\mu} D^{-1}(g) \cdot f_l^{\kappa,\nu\dagger}(g^{-1}\bm{k}) \\
&=\sum_{\kappa,\mu,\nu}a_{\kappa \mu\nu}\vec{X}^{\kappa ,\mu}D^\kappa (g)
\cdot (\vec{f}^{\kappa,\nu}(\bm{k})D^\kappa (g))^\dagger \\
&=\sum_{\kappa,\mu,\nu}a_{\kappa \mu\nu} \vec{X}^{\kappa, \mu} \cdot 
D^\kappa (g)D^{\kappa \dagger}(g)
\cdot \vec{f}^{\kappa,\nu\dagger }(\bm{k})\\
&=\sum_{\kappa,\mu,\nu}a_{\kappa \mu\nu}\vec{X}^{\kappa,\mu} \vec{f}^{\kappa,\nu\dagger}(\bm{k})\ \ \ (D^\kappa \  \text{unitary})\\
&=H(\bm{k})
\end{aligned}
\end{equation}
Anti-unitary symmetries can be proved similarly.
The unitary condition of coirreps is easily satisfied, as any linear representation of a finite group is equivalent to a unitary representation. 
It is worth mentioning that when matching $\vec{X}^{\kappa,\mu}$ and $\vec{f}^{\kappa,\nu}(\bm{k})$, not only the coirrep, but also the corepresentation matrices of anti-unitary symmetries need to be identical (which in general could differ by a phase factor for equivalent coirreps). Moreover, to ensure the Hermiticity of the final Hamiltonian, if some matched term $\vec{X}^{\kappa,\mu}\vec{f}^{\kappa,\nu\dagger}(\bm{k})$ is not Hermitian, it should 
be combined with its conjugated term, i.e., 
$\vec{X}^{\kappa,\mu}\vec{f}^{\kappa,\nu\dagger}(\bm{k})+\vec{X}^{\kappa,\mu'}\vec{f}^{\kappa,\nu'\dagger}(\bm{k})$, where $\mu',\nu'$ are defined by taking complex conjugation to the similarity transformation matrices: $U^{\kappa,\mu'}=U^{\kappa,\mu*}, V^{\kappa,\nu'}=V^{\kappa,\nu*}$.

We remark that the two algorithms are equivalent. In the direct-product decomposition method, assume we have taken proper basis of $\vec{f}(\bm{k})$ and $\vec{X}$ s.t. the reducible representation matrices $N(\hat{F}_g)$ and $M(\hat{G}_g)$ are block diagonal, with each block corresponds to some linear coirrep $D^\kappa$.  
Using the following theorem in group representation theory, it can be shown that only the direct product of two conjugated coirreps can be decomposed into a trivial coirrep, which survives in the final Hamiltonian:
\begin{theorem}
	The decomposition of the direct product of two (co)irreps $D^\kappa\otimes D^\lambda$ contains the trivial (co)irrep $\iff$ $D^\kappa$ is equivalent to $D^{\lambda*}$.
\end{theorem}
As a result, each independent term in the Hamiltonian corresponds to a linear coirrep $D^\kappa$ and has the form $\vec{X}^{\kappa,\mu}\vec{f}^{\kappa ,\nu\dagger}(\bm{k})$.

\section{The kdotp-generator package}

The \textit{kdotp-symmetry} package\cite{gresch2018identifying} developed by D. Gresch is written in \textit{Python} and makes use of the \textit{SymPy} library for main symbolic computation. This package takes in the $O(3)$ rotation and representation matrices of (magnetic) little group generators as well as a given order of $\bm{k}$, and outputs all compatible $\bm{k}\cdot\bm{p}$ models.

Based on the \textit{kdotp-symmetry} package, we make the following improvements and develop an upgraded package \textit{kdotp-generator}:
\begin{itemize}
	\item Generalize the input so that it can compute the effective Hamiltonians of external fields and their couplings to $\bm{k}$.
	\item Add a post-processing step to decompose the Hamiltonian into symmetrical monomial function and Hermitian matrix bases using linear coirreps.
	\item Pre-compute effective Hamiltonians in MSGs up to the third order.
	\item Slightly improve the efficiency of the code.
\end{itemize}

To use this package, users can either input their own (reducible or irreducible) representation matrices and compute the $\bm{k}\cdot\bm{p}$ models, or directly refer to our pre-computed results. Details of the package are given on the website \href{https://github.com/yjiang-iop/kdotp-generator}{https://github.com/yjiang-iop/kdotp-generator}.

In the pre-computation step, we exhaust effective Hamiltonians for coirreps in MSGs by first tabulating (both single-valued and double-valued, projective) coirreps for each high-symmetry momenta of 1651 MSGs. To achieve this, we make use of the irreps of 230 double space groups and the symmetry operations of MSGs from the Bilbao website\cite{aroyo2006bilbao1, aroyo2006bilbao2, aroyo2011crystallography}  and follow the standard method\cite{bradley2009mathematical} to derive the coirreps of each high-symmetry momenta from the irreps of the unitary halving subgroup of the corresponding magnetic little group\footnote{Recently, Bilbao website has updated the coirreps of 1651 MSGs\cite{elcoro2020magnetic}. Nonetheless, we use our homemade code to generate the coirreps of MSGs, which are not exactly the same as those on Bilbao website, especially for some type-3 MSGs.}. 
For each magnetic little group, we also compute its linear coirreps by identifying its corresponding magnetic point group, which is equivalent to the magnetic little group of $\Gamma$ of the corresponding symmorphic MSG. Then the coirreps of symmorphic MSGs are used to generate linear coirreps.

With all projective and linear coirreps of MSGs derived, we feed them into the \textit{kdotp-generator} package to obtain all symmetry-allowed independent $\bm{k}\cdot\bm{p}$ models up to the third order and decompose them using linear coirreps. We also calculate the Hamiltonian of the electromagnetic field $\bm{E},\bm{B}$ and the strain tensor $\epsilon_{\mu\nu}$ up to the third order, as well as their linear couplings to $\bm{k}$. These results are included in the \textit{kdotp-generator} package. Users can also input reducible representations consist of multiple coirreps and calculate the corresponding models.

We remark that the direct-product decomposition method can compute the effective Hamiltonians without the pre-knowledge of linear coirreps of the magnetic little group, which is the main advantage of this algorithm. However, if the coirreps are available, the algorithm can be simplified by directly decomposing the representation matrices $M(\hat{G}_g)$ and $N(\hat{F}_g)$ using the coirreps which gives symmetric basis $\vec{X}^{\kappa,\mu}$ and $\vec{f}^{\kappa,\nu}$, and then using the irrep-matching method to obtain the results.
Nonetheless, we follow the direct-product decomposition method and decompose the output Hamiltonians as a post-processing step.

\section{Example}
We use type-3 MSG 10.44 $P2'/m$ as an example to show the output of the package. The high-symmetry momenta $\Gamma$ has magnetic little group $2'/m$, which has two generators:
\begin{equation}
\begin{aligned}
M_{y}&=
\left(\begin{array}{ccc}
1 & 0 & 0  \\
0 & -1 & 0   \\
0 & 0 & 1  \\
\end{array}\right)\\
C_{2y}\cdot T&=
\left(\begin{array}{ccc}
-1 & 0 & 0  \\
0 & 1 & 0   \\
0 & 0 & -1  \\
\end{array}\right)\cdot \hat{T}\\
\end{aligned}
\end{equation}
We choose a two-dimensional coirrep $\overline{\Gamma}_3\overline{\Gamma}_4$, whose representation matrices are
\begin{equation}
\begin{aligned}
D(M_{y})&=
\left(\begin{array}{cc}
-i & 0 \\
0 & i   \\
\end{array}\right)\\
D(C_{2y}\cdot T)&=
\left(\begin{array}{cc}
0 & 1 \\
1 & 0   \\
\end{array}\right)\cdot \hat{K}\\
\end{aligned}
\end{equation}

The computed independent effective Hamiltonians of $\bm{k},\bm{E},\bm{B}$ and their linear couplings are listed in Table.\ref{table_kEB},\ref{linear_couple}.

\begin{table}[H]
	\centering
	\begin{tabular}{c|c|c|c}
		\hline\hline
		Variable & 0th order & 1st order & 2nd order \\\hline
		$\bm{k}$ & $\sigma_0$ & 
		$k_x\sigma_0, k_z\sigma_0$ &
		\makecell[c]{$k_x^2\sigma_0, k_y^2\sigma_0$,\\ $k_z^2\sigma_0, k_xk_z\sigma_0$} \\\hline
		$\bm{E}$ & $\sigma_0$ & 
		\makecell[c]{$E_x\sigma_z, E_y\sigma_x,$\\$E_y\sigma_y, E_z\sigma_z$} &
		\makecell[c]{$E_x^2\sigma_0, E_y^2\sigma_0$,\\ $E_z^2\sigma_0, E_xE_z\sigma_0$} \\\hline
		$\bm{B}$ & $\sigma_0$ & 
		\makecell[c]{$B_x\sigma_x, B_x\sigma_y, B_y\sigma_z$,\\ $B_z\sigma_x, B_z\sigma_y$} & 
		\makecell[c]{$B_x^2\sigma_0, B_y^2\sigma_0$,\\ $B_z^2\sigma_0, B_xB_z\sigma_0$}  \\\hline\hline
	\end{tabular}
	\caption{\label{table_kEB} Independent Hamiltonians of $\bm{k},\bm{E},\bm{B}$ up to the second order of MSG 10.44, coirrep $\overline{\Gamma}_3\overline{\Gamma}_4$.}
\end{table}

\begin{table}[ht]
	\centering
	\begin{tabular}{c|c}
		\hline\hline
		Variable & Linear coupling terms \\\hline
		$\bm{k},\bm{E}$ & \makecell[c]{
		$E_xk_x\sigma_z, E_yk_x\sigma_x, E_yk_x\sigma_y, E_zk_x\sigma_z, E_xk_y\sigma_x$,\\ 
		$E_xk_y\sigma_y, E_yk_y\sigma_z, E_zk_y\sigma_x, E_zk_y\sigma_y, E_xk_z\sigma_z$,\\ 
		$E_yk_z\sigma_x, E_yk_z\sigma_y, E_zk_z\sigma_z$}
		 \\\hline
		 $\bm{k},\bm{B}$ & \makecell[c]{
		 $B_xk_x\sigma_x, B_xk_x\sigma_y, B_yk_x\sigma_z, B_zk_x\sigma_x, B_zk_x\sigma_y$,\\
		 $B_xk_y\sigma_z, B_yk_y\sigma_x, B_yk_y\sigma_y, B_zk_y\sigma_z, B_xk_z\sigma_x$,\\
		 $B_xk_z\sigma_y, B_yk_z\sigma_z, B_zk_z\sigma_x, B_zk_z\sigma_y$} 
		\\\hline
		$\bm{E},\bm{B}$ & 
		$B_yE_x\sigma_0, B_xE_y\sigma_0, B_zE_y\sigma_0, B_yE_z\sigma_0$
		\\\hline\hline
	\end{tabular}
	\caption{\label{linear_couple}Independent Hamiltonians of the linear couplings of $\bm{k},\bm{E},\bm{B}$ of MSG 10.44, coirrep $\overline{\Gamma}_3\overline{\Gamma}_4$.}
\end{table}

As another example, we use type-1 MSG 195.1 $P23$ to show the post-processing step of decomposing the $\bm{k}\cdot\bm{p}$ models into symmetric bases of $\bm{k}$-monomials and Hermitian matrices. The irrep $\overline{\Gamma}_5$ of $\Gamma$ has two generators $C_{2z},C_{3,111}^+$, whose rotation and representation matrices are 
\begin{equation}
\begin{aligned}
C_{2z}&=
\left(\begin{array}{ccc}
-1 & 0 & 0  \\
0 & -1 & 0   \\
0 & 0 & 1  \\
\end{array}\right),\
D(C_{2z})=
\left(\begin{array}{cc}
-i & 0 \\
0 & i   \\
\end{array}\right)\\
C_{3,111}^+&=
\left(\begin{array}{ccc}
0 & 0 & 1  \\
1 & 0 & 0   \\
0 & 1 & 0  \\
\end{array}\right),\
D(C_{3,111}^+)=\frac{1}{2}
\left(\begin{array}{cc}
1-i & -1-i \\
1-i & 1+i   \\
\end{array}\right)\\
\end{aligned}
\end{equation}
The decomposed $\bm{k}\cdot\bm{p}$ models are listed in Table.\ref{table_decompose}.
\begin{table}[ht]
	\centering
	\begin{tabular}{c|c|c|c|c}
		\hline\hline
		Order & 0th  & 1st  & \multicolumn{2}{c}{2nd}  \\\hline
		$\vec{f}(\bm{k})$ & $1$ & $(k_x,k_y,k_z)$ & $k_x^2+k_y^2+k_z^2$ 
		& $(k_yk_z, k_xk_z, k_xk_y)$ \\\hline
		$\vec{X}$ & $\sigma_0$ & $(\sigma_x,\sigma_y,\sigma_z)$ & $\sigma_0$ &
		$(\sigma_x,\sigma_y,\sigma_z)$
		\\\hline\hline
	\end{tabular}
	\caption{\label{table_decompose}Decomposed $\bm{k}\cdot\bm{p}$ models of MSG 195.1 $\overline{\Gamma}_5$. Notice there are two independent terms in the 2nd order.}
\end{table}

\textit{Note added}: In the final stage of this work, we notice two similar works\cite{yu2021encyclopedia, tang2021exhaustive} which also calculated and analyzed the $\bm{k}\cdot\bm{p}$ effective Hamiltonians for SGs and MSGs.

\bibliography{ref}

\end{document}